\documentclass[prl, twocolumn, showpacs]{revtex4}
\usepackage{amsmath, amsfonts, amsthm, bbm,bm}
\usepackage{natbib} 
\usepackage{color}
\usepackage[latin1]{inputenc}
\usepackage[final]{graphicx}

\newcommand{\mb}[1]{\text{\boldmath ${#1}$}}

\newcommand{\abs}[1]{\left\lvert#1\right\rvert}
\newcommand{\abss}[1]{\lvert#1\rvert}
\renewcommand{\Im}{\operatorname{Im}} 
\renewcommand{\Re}{\operatorname{Re}} 
\newcommand{\Pint}{\mathcal{P}\negthickspace\negthickspace\int} 
\newcommand{\Tr}{\operatorname{Tr}} 
\def\ket#1{\mathinner{|{#1}\rangle}}
\renewcommand{\[}{\begin{equation}}
\renewcommand{\]}{\end{equation}}

\newcommand{\ITPFUB}{Institut f\"ur Theoretische Physik, Freie Universit\"at Berlin, Fachbereich Physik, 14195 Berlin, Germany}

\newcommand{\kBT}{k_\text{B} T}
\renewcommand{\L}{\mathcal{L}}
\renewcommand{\P}{\mathcal{P}}
\newcommand{\Q}{\mathcal{Q}}
\newcommand{\R}{\mathbb{R}}

\begin{document}
\title{Quantum Transport through Nanostructures in the Singular-Coupling Limit}
\author{Maximilian G.~Schultz}
\author{Felix von Oppen}
\affiliation{\ITPFUB}
\date{\today}
\begin{abstract} 

  Geometric symmetries cause orbital degeneracies in a molecule's spectrum. In a single-molecule junction, these
  degeneracies are lifted by various symmetry-breaking effects. We study quantum transport through such nanostructures
  with an almost degenerate spectrum. We show that the master equation for the reduced density matrix must be derived
  within the singular-coupling limit as opposed to the conventional weak-coupling limit. This results in strong
  signatures of the density matrix's off-diagonal elements in the transport characteristics.
  
\end{abstract} 
\maketitle

\textit{Introduction.}---The Anderson model of localized, interacting electronic states coupled to two Fermi-gas
electrodes is the archetypical model for studying electronic transport through quantum nanostructures. The model
accurately describes a wide range of experimentally accessible transport regimes of quantum dots including the
Coulomb blockade and the Kondo regime. It has been successfully extended to describe single-molecule junctions
by coupling the electrons to the vibrational degrees of freedom of the molecule. The study of molecular electronics
using this model has received enormous attention in recent years \cite{Galperin07}.

An important issue that in contrast to the vibrational structure has been studied much less arises from the orbital
degeneracies due to the geometric symmetries of molecules. The transport properties due to orbital degeneracies differ
from those due to spin degeneracy, which are being studied extensively in the quantum-dot literature. Since there, the
underlying $SU(2)$ symmetry is respected both by the electrons in the reservoirs and by the tunnel couplings to the
localized system, the low-temperature phenomenology is characterized by Kondo physics \cite{Pustilnik05}.

The symmetry-induced orbital degeneracies of a molecule are in general lifted by binding it to metallic electrodes or
the interaction with an underlying substrate. Asymmetries in the coupling of the orbitals to the leads cause a
tunneling-induced splitting of the degenerate levels as the result of perturbation theory.  The problem with orbital
degeneracies therefore naturally extends to studying the role of near-degeneracies in quantum transport theory.

The importance of coherent superpositions of degenerate localized levels for quantum transport has already been
mentioned in the literature \cite{Braun04, Darau08}.  In this paper, we generalize the discussion from the non-generic
case of exact degeneracy and show how the breaking of symmetries and the lifting of molecular degeneracies can be
consistently accounted for in a master equation formalism by employing the ``singular-coupling limit'' \cite{Spohn80} in
the derivation of the kinetic equation for sequential tunneling.  Contrary to the weak-coupling limit, the
singular-coupling limit can properly describe the coherent dynamics in the near-degenerate orbital subspace of the
reduced density matrix and its competition with the transport dynamics due to electron tunneling.  Our approach shows
that this competition causes rich and interesting physics in the transport characteristics.

The singular-coupling limit also remedies the shortcomings of the Bloch--Redfield equation, which is a master equation
that explicitly keeps the coherences between non-degenerate states in the density matrix \cite{Timm08} but is known to
produce negative probabilities \cite{Duemcke79}.  The master equation in the singular-coupling limit is included in the
general theoretical framework of master equations and rate equations. It closes the gap between the description by rate
equations and degenerate master equations enabling us to study high-temperature sequential tunneling for all possible
energy regimes.

We illustrate the physics of an orbital near-degeneracy within a minimal model, a two-level, interacting but spinless
Anderson impurity coupled to two electrodes
\[\begin{split}\label{eq:Anderson}
  H = & eV_\text{g} (n_\uparrow + n_\downarrow) + \frac{\Omega}{2}(n_\uparrow - n_\downarrow) + U n_\uparrow n_\downarrow \\
& + \sum_{\mb{k}\alpha} \varepsilon_\mb{k} c^\dagger_{\mb{k}\alpha} c_{\mb{k}\alpha}  + \sum_{\mb{k}\alpha\sigma}
t_{\alpha \sigma} c^\dagger_{\mb{k}\alpha}d_\sigma + \text{h.c.}
\end{split}
\]
The on-site electronic orbitals are labeled by a pseudo-spin $\sigma = \uparrow,\downarrow$ and are separated in energy
by $\Omega$. This splitting is assumed to be due to symmetry-breaking mechanisms other than electronic tunneling. Double
occupation of the system is suppressed by Coulomb repulsion of strength $U$, and both orbitals are coupled to Fermi-gas
electrodes $\alpha = L,R$, held in thermal equilibrium at temperature $\kBT$, via lead- and orbital-dependent amplitudes
$t_{\alpha\sigma}$.

\begin{figure}
 \begin{center}
  \begin{minipage}[c]{\linewidth}
    \includegraphics[width=\linewidth]{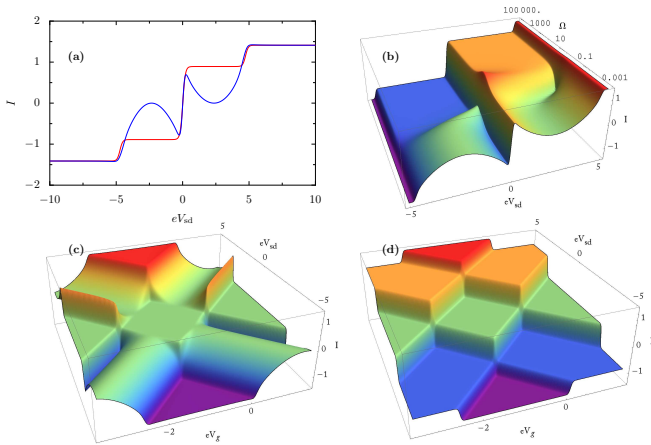}
  \end{minipage}
  \caption{(Color online) Numerical evaluation of the stationary current Eq.~(\ref{eq:pseudo-curr}) through an
  interacting two-level Anderson model, $t_{\text{L}\uparrow} = t_{\text{R}\downarrow} = 1$, $t_{\text{L}\downarrow} =
  t_{\text{R}\uparrow} = 1.5$, $U = 2.35$, $\kBT = 0.02$. The tunnel amplitudes $t_{\alpha\sigma}$ and the splitting
  $\Omega$ are measured in units of $\frac{2\pi}{\hbar}\nu_0$. Fig.~(a) illustrates the difference between the
  description by the degenerate master equation, Eqs.~(\ref{eq:pseudo-Bloch}, \ref{eq:master_pop}), (blue) and the $\Omega = 0$ rate
  equation, which is obtained from the master equation by setting $\rho_{\uparrow\downarrow}=0$ (red). Both models are
  evaluated at $eV_\text{g} = 0$ in order to highlight the quantitative as well as qualitative discrepancy. In Figs.~(c) and
  (d), we show a full scan of the $(eV_\text{g}, eV_\text{sd})$-plane for both equations, illustrating the fundamental
  difference of the dynamics due to the two equations in the parameter space. While the rate-equation result is the
  familiar sequence of steps in the stationary current, the master-equation result shows a strong current suppression
  for voltages below the double-charging threshold. Using the master equation in the singular-coupling limit as it is
  derived in the text, we interpolate the two different curves of Fig.~(a) by letting $\Omega\to \infty$ in
  Fig.~(b).\label{fig:SC-cross-over}}
 \end{center}
\end{figure}

In Fig.~\ref{fig:SC-cross-over}~(a) we show the stationary current through the model system for $eV_\text{g} = 0$. It is
evaluated both for vanishing $\Omega$ using a master equation that includes the off-diagonal elements of the reduced
density matrix $\rho$, and for finite $\Omega$, which is small compared with $\kBT$, using a rate equation for the
diagonal elements of $\rho$. The \textit{qualitative} difference between the two results is striking. While the rate
equation yields a simple steplike increase of the current, the master equation produces additional structure: a
suppression of the current and pronounced negative differential conductance. The phenomenological discrepancy between
the two curves of Fig.~\ref{fig:SC-cross-over}~(a) illustrates that there can be interesting and non-intuitive physics in
the cross-over regime where $\Omega$ is of the order of the tunneling-induced broadening $\Gamma$ of the electronic
orbitals.

\textit{The Singular-Coupling Limit.}---The emergence of additional physics in the cross-over regime $\Omega \sim \Gamma$ can
already be understood with the help of the non-interacting, $U=0$, version of the two-level Anderson model,
Eq.~(\ref{eq:Anderson}).  Due to the absence of interactions, the retarded propagator of the model can be computed
exactly, and a straight-forward calculation yields for the off-diagonal element of the spectral function
\[
\mathcal{A}_{\uparrow\downarrow}(\omega) =
\frac{\gamma\left(\left(\omega-\frac{\Omega}{2}\right)\left(\omega+\frac{\Omega}{2}\right) - \frac{\Delta}{4}
\right)}{\abs{\left(\omega-\frac{\Omega}{2}+\imath\frac{\Gamma_\uparrow}{2}\right)\left(\omega+\frac{\Omega}{2}+\imath\frac{\Gamma_\downarrow}{2}\right) + \frac{\gamma^2}{4}}^2}.
\]
The self energy $\Sigma = \imath \frac{\pi}{\hbar}\nu_0 W^\dagger W$, where $(W)_{\alpha\sigma}:=t_{\alpha\sigma}$, gives
rise to the broadenings $\Gamma_{\alpha\sigma}:= \frac{2\pi}{\hbar}\nu_0t_{\alpha\sigma}^2$  and $\gamma_\alpha :=
\frac{2\pi}{\hbar}\nu_0 t_{\alpha\uparrow}t_{\alpha\downarrow}$,  with a missing index indicating that it has been
summed over; the density of states $\nu_0$ in the electrodes is assumed to be uniform. We also set
$\delta\Gamma:=\Gamma_\uparrow - \Gamma_\downarrow$. The parameter
$\Delta:=\Gamma_\uparrow\Gamma_\downarrow - \gamma^2$ is proportional to $\det(\Sigma)$. A vanishing $\Delta$ allows for
a complete decoupling of one electronic level from both electrodes by a unitary transformation of the Hamiltonian
\cite{ShahbRaikh94,Berkovits04}. We therefore focus on the generic situation $\Delta \neq 0$ in the following.

The stationary coherences of the non-degenerate system are described by the equal-time correlator $\int_\R
\mathcal{A}_{\uparrow\downarrow}(\omega)\, d \omega$. In the weak-coupling limit, when $\Gamma \to 0$, these are
expected to vanish for every finite $\Omega$. To confirm this explicitly, we rescale $t_{\alpha\sigma} \mapsto \lambda
t_{\alpha\sigma}$ and shift the energy $\xi:=\omega - \frac{\Omega}{2}$. As $\lambda$ tends to zero, the rescaled
spectral function $\mathcal{A}^\lambda_{\uparrow\downarrow}(\xi)$ converges to zero almost everywhere except at the four
roots of its denominator,
\[\label{eq:roots}
\xi_{1,2}  =  \frac{\lambda^2}{4}\left[\delta\Gamma - 2\frac{\Omega}{\lambda^2} -\imath\Gamma \pm
  2\sqrt{\left(\frac{\Omega}{\lambda^2} \right)^2 -  \gamma^2}\,\, \right]
\]
and $\xi_{3,4} = \xi_{1,2}^\ast$. By use of the residue theorem, we find that also the contribution of these poles to the
integral converges to zero such that $\int_\R A_{\uparrow\downarrow}^\lambda(\xi)\,d\xi \to 0$ as $\lambda \to 0$.
When, on the contrary, both $\Gamma$ and $\Omega$ are of the same order, we have to rescale not only $t_{\alpha\sigma}$
but also $\Omega \mapsto \lambda^2\Omega$, such that $\Gamma_{\alpha\sigma}/\Omega$ and $\gamma_\alpha/\Omega$ remain
fixed in the limit process. This simultaneous scaling is known as ``singular-coupling limit'' in contrast to the
weak-coupling limit, in which $\Omega$ is kept constant \cite{Spohn80}. Repeating the above calculation, the roots of the
spectral function's denominator are now proportional to $\lambda^2$, and the integral
\[
\int_\R \mathcal{A}_{\uparrow\downarrow}^\lambda(\xi)\,d\xi = 2\pi\frac{2\gamma\delta\Gamma^2}{\Gamma\bigl(\Gamma^2 +
4(\Omega^2 - \gamma^2)\bigr)}
\]
is independent of the scaling parameter and finite. Although the system is non-degenerate, some coherences between the
electronic levels are retained. As we let $\Omega \to \infty$, these converge to zero. In this limit, our approach
reproduces the weak-coupling result.

\textit{Master Equation.}---We now derive the master equation for the near-degenerate, \textit{interacting} two-level
Anderson model in the singular-coupling limit. The underlying challenge of the regime $\Omega \sim \Gamma$ is that the
tunneling dynamics takes place on the very same time scale as the coherent on-site dynamics due to the almost degenerate
electronic levels. We consider the  Hamiltonian, rescaled according to the previous observation, $H^\lambda = \lambda^2
H_\text{S} + H_\text{E} + \lambda H_\text{S--E}$.  The term $\lambda^2 H_\text{S}$ describes the system, that is the
near-degenerate electronic orbitals. We set $eV_\text{g} = 0$ for convenience---it can easily be restored later on.
$H_\text{E}$ is the bath, in our case the Fermi-gas electrodes, and $\lambda H_\text{S--E}$ is the system--bath
interaction, the tunnel Hamiltonian. We rewrite the von Neumann equation as an integral equation for the reduced density
matrix $\rho_\text{S} = \Tr_\text{E}(\rho)$.  The limit $\lambda \to 0$ then generates the markovian dynamics
\cite{Davies74}. The reduced density matrix itself is obtained by a projection $\P\rho := \Tr_\text{E}(\rho) \otimes
\rho_\text{E}$, with $\rho_\text{E}$ being the equilibrium distribution of the bath. With the complementary projection
$\Q := 1-\P$, the von Neumann equation $\dot{\rho} = -\imath [H^\lambda, \rho] = -\imath\L^\lambda\rho =
-\imath(\lambda^2 \L_\text{S} + \L_\text{E} + \lambda\L_\text{S--E})\rho$ reads in the interaction picture
\cite{Duemcke79,Davies74}
\begin{widetext}
\[\label{eq:Master_Integral}
\rho^\text{I}_\text{S}(t) = \rho^\text{I}_\text{S}(0) - \lambda^2 \int_0^t
e^{\imath\lambda^2\L_\text{S}v}\left\{\int_0^{t-v} e^{\imath\lambda^2\L_\text{S}s}
\Tr_\text{E}\left(\L_\text{S--E}\Q e^{-\imath\L^\lambda s}\Q \L_\text{S--E}
\rho_\text{E}\right)\, d s \right\}e^{-\imath\lambda^2\L_\text{S}v}\,\rho^\text{I}_\text{S}(s)\,d v.
\]
\end{widetext}
This equation incorporates the Born approximation by choosing $\rho(0) = \rho_S(0)\otimes\rho_\text{E}$ as the initial
condition. On the slow markovian time scale $\tau = \lambda^2 t$, the term in curly brackets converges to a time-independent
quantity \cite{Davies74}: the master equation assumes the markovian form 
\[\label{eq:SC-Master}
\dot{\rho}_\text{S} = -\imath\L_\text{S}\rho_\text{S} - \int_0^\infty \!\!\!\! \Tr_\text{E}\left(\L_\text{S--E}e^{-\imath\L_\text{E}s}\L_\text{S--E}\rho_\text{E} \right)d s\,\rho_\text{S}.
\]
If $H_\text{S}$ had not been rescaled by $\lambda^2$, the exponential factors enclosing the curly bracket in
Eq.~(\ref{eq:Master_Integral}) due to the interaction picture would have taken the form
$\exp(\pm\imath \L_\text{S} \tau/\lambda^2)$ on the markovian time scale $\tau$. In this case, letting $\lambda\to 0$
would have caused faster and faster oscillations for coherences belonging to non-degenerate states. In the limit, those
terms would have been averaged to zero, commonly known as the secular approximation \cite{Breuer02}. 

As it is plausible from Eq.~(\ref{eq:SC-Master}), the master equation in the singular-coupling limit is formally
obtained by setting $\Omega=0$ in the dissipative term but retaining the non-zero $\Omega$ in the free-evolution
Hamiltonian \cite{Darau08}. The splitting $\Omega$ must not appear in the Fermi functions coming from the trace over the bath degrees
of freedom, because due to the limit $\lambda\to 0$, the temperature of the bath is too large to resolve the splitting.
The Bloch--Redfield equation would, however, have such a dependence on $\Omega$.

We now evaluate the master equation (\ref{eq:SC-Master}) for the Anderson-model Eq.~(\ref{eq:Anderson}) and recast it as
a Bloch equation for the pseudo-spin $\vec{S} := (2\Re \rho_{\uparrow\downarrow}, 2\Im\rho_{\uparrow\downarrow},
\rho_{\uparrow\uparrow} - \rho_{\downarrow\downarrow})$, which is defined by the electronic two-level system, coupled to an
equation for the populations $p_i$, $i$ being the number of electrons on the device,
\begin{widetext} \begin{align}  \dot{\vec{S}} & =  \sum_\alpha\left[f_{\alpha}p_0 +
  \frac{1}{2}\Bigl(f_{\alpha 2} - (1-f_{\alpha}) \Bigr)p_1 - (1-f_{\alpha 2})p_2
    \right]\vec{n}_\alpha - \frac{1}{2}\sum_\alpha\Bigl[f_{\alpha 2} + (1-f_{\alpha}) \Bigr]\Gamma_\alpha\vec{S} -
    (\vec{B} + \Omega \, \hat{e}_z)\times \vec{S}\label{eq:pseudo-Bloch}, \\
    \frac{d}{dt}\begin{pmatrix}p_0\\ p_1 \\ p_2\end{pmatrix} & = \frac{1}{2} \sum_\alpha 
    \Gamma_\alpha\begin{pmatrix}
      -2f_{\alpha} & (1-f_{\alpha}) & 0 \\
      2f_{\alpha} & - f_{\alpha 2} - (1-f_{\alpha}) & 2(1-f_{\alpha 2})\\
      0 & f_{\alpha 2} & - 2(1-f_{\alpha 2})
    \end{pmatrix}
    \begin{pmatrix} p_0 \\ p_1 \\ p_2\end{pmatrix}
      + \frac{1}{2} \sum_\alpha \begin{pmatrix} 1-f_{\alpha}\\ f_{\alpha 2} - (1-f_{\alpha}) \\ -f_{\alpha 2} \end{pmatrix} \vec{n}_\alpha
	\cdot\vec{S},\label{eq:master_pop}\\[\medskipamount]
	I_\alpha & = \Gamma_\alpha\left[ f_\alpha p_0 -\Bigl((1-f_\alpha) - f_{\alpha 2}\Bigr) p_1  -
	(1-f_{\alpha 2})p_2 \right] - \frac{1}{2} \Bigl((1-f_\alpha) + f_{\alpha 2} \Bigr) \vec{n}_\alpha\cdot\vec{S}.\label{eq:pseudo-curr}
      \end{align}
\end{widetext}
Eq.~(\ref{eq:pseudo-curr}) gives the stationary current through lead $\alpha$ by $I_\alpha = \Tr_\text{S}(\rho
\hat{I}_\alpha)$.The pseudo-magnetic fields are determined by the principal-value integrals
\[\label{eq:B_field}
\vec{B}  :=  \frac{1}{2\pi}\sum_\alpha\Pint f_\alpha(\varepsilon)\left(\frac{1}{U - \varepsilon} +
\frac{1}{\varepsilon}\right)d\varepsilon\,\vec{n}_\alpha,
\]
with $\vec{n}_\alpha := (2\gamma_\alpha, 0, \Gamma_{\alpha\uparrow} - \Gamma_{\alpha\downarrow})$. The Fermi functions
are abbreviated by $f_\alpha := f(eV_\text{g}-\mu_\alpha)$ and $f_{\alpha 2}:= f(eV_\text{g} + U-\mu_\alpha)$.
For $\Omega = 0$, Eqs.~(\ref{eq:pseudo-Bloch}--\ref{eq:pseudo-curr}) are reminiscent of the ones derived in
\cite{Braun04}.

In Eqs.~(\ref{eq:pseudo-Bloch}, \ref{eq:master_pop}), the electronic splitting term $-\imath\L_\text{S}\rho_\text{S}$
appears as a contribution to the pseudo-magnetic field directed along $\hat{e}_z$. Due to this interpretation, the
pseudo-magnetic fields describe the tunneling-induced energy renormalizations of the electronic levels in exactly the
same way as any other term in the Hamiltonian of this order would have to be treated, namely in the singular-coupling
limit. The singular-coupling limit provides a non-discriminating description for both, the tunneling-induced
renormalization of the on-site levels as the \textit{result} of perturbation theory and any other splitting of the same
order, which has to be included in the model \textit{explicitly}.

\textit{Results.}---Based on our theoretical considerations, we understand the qualitative behavior of the stationary
current--voltage characteristics of the two-level Anderson model and its remarkable sensitivity to $\Omega$. The
rate-equation results of Figs.~\ref{fig:SC-cross-over} (a) and (d) show the familiar steps in the stationary current
whenever the voltage is large enough to add another electron to the device. Since $\Omega \ll \kBT$, the near-degeneracy
remains unresolved.

The $\Omega = 0$ master equation, also yields a finite linear conductance at zero bias, but below the double-charging
threshold, in Figs.~\ref{fig:SC-cross-over}~(a) and (c), the stationary current is strongly suppressed.  Since the
equation is one for the full density matrix of the system's on-site levels, the underlying mechanism is explained by
choosing a particular electronic basis. The key idea is: for general tunneling amplitudes $t_{\alpha\sigma}$, there
always exists a unitary transformation to eliminate at least one of them and hence decouple one of the levels from one
electrode. Let this level be $\ket{\uparrow}$ and let the electrode it is decoupled from be the drain electrode,
$\Gamma_{\text{d}\uparrow} = 0$.  Then $\gamma_\text{d} = 0$.  If we neglect the pseudo-magnetic fields for the time
being, the degenerate master equation is actually only a rate equation, whose physics is easily understood.  The
tunneling dynamics will eventually populate the decoupled state $\ket{\uparrow}$, which due to
$\Gamma_{\text{d}\uparrow} = 0$ cannot be left again. Below the double-charging threshold Coulomb repulsion then
obstructs any current through the device.

The pseudo-magnetic fields, which describe virtual switching processes between the degenerate levels, soften this
picture.  They induce a precession of the pseudo-spin that corresponds to moving the electron from the decoupled state
$\ket{\uparrow}$ via virtual intermediate states on the source electrode into the conducting state $\ket{\downarrow}$
\cite{Braun04}.  Except for those voltages, for which $\abss{\vec{B}} = 0$ and the above argument for complete current
suppression applies, the device carries the highly voltage-dependent stationary current shown in
Figs.~\ref{fig:SC-cross-over}~(a) and (d).

In the singular-coupling limit, the splitting $\Omega$ generates an additional pseudo-magnetic field of the same order
as the tunneling-induced one. Its orientation along $\hat{e}_z$ distinguishes this direction in the system's Hilbert
space. As $\Omega$ is being increased, the precession frequency of $\vec{S}$ about the $z$-axis will dominate the
virtual tunneling dynamics, which are of the order of the electronic dwell time on the device. In the limit $\Omega \to
\infty$, the residual precession dynamics is too fast compared with the average tunneling time, such that only the
projection of $\vec{S}$ onto the $z$-axis is relevant for the tunneling dynamics. For very large $\Omega$, the
pseudo-spin is effectively oriented along $\hat{e}_z$. Because then $S_x + \imath S_y = 2\rho_{\uparrow\downarrow}
\approx 0$, the master equation is rendered a rate equation for the electronic occupations. As still $\Omega \ll \kBT$,
the Fermi functions cannot resolve the physics due to $\Omega$, and the system is effectively described by the $\Omega =
0$ rate equation.

The numerical results shown in Fig.~\ref{fig:SC-cross-over}~(b) support this reasoning: as the splitting $\Omega$ is
numerically increased, the current-suppression curve is shifted toward positive bias. The suppression of the stationary
current is lifted and eventually lost in the flat current profile of the $\Omega = 0$ rate-equation result.

\textit{Conclusions.}---The omnipresence of orbital degeneracies in molecular physics and their lifting due to various
symmetry-breaking effects in single-molecule junctions requires the modeling of degenerate and near-degenerate systems
in quantum transport theory.

We have shown that within the framework of master equations for sequential tunneling, such near-degenerate systems fall
into a descriptive gap between the energy regimes that are discussed in the theoretical literature. Whereas for
degenerate systems, a master equation for the full reduced density matrix is used, systems with energy differences
larger than the tunneling-induced broadening $\Gamma$ have to be described by rate equations. In the interesting
cross-over regime, the tunneling of electrons, the tunneling-induced splitting, and the coherent on-site dynamics are
both of order $\Gamma$.  By using the notion of the singular-coupling limit, we have derived a master equation for this
regime. We have thereby also given meaning to the rate-equation treatment of degenerate systems as being the proper
description when $\Gamma \ll \Omega \ll \kBT$. Since for the high-temperature regime always $\Gamma \ll \kBT$ is
implied, we have actually presented a method to describe all possible energy regimes for sequential tunneling through
multilevel quantum nanostructures.

Aside from the simple model that we have used to illustrate and explain its physics, our approach has a wide range of
applications in molecular electronics. It provides a means to generically account for symmetry-breaking mechanisms in
single-molecule junctions. And it allows the study of complex molecular models such as Jahn--Teller active systems,
pseudo-Jahn--Teller structures,  the valley degeneracy in carbon nanotubes, or the interaction of
orbital symmetries and vibrational degrees of freedom, which is a leitmotif in the theory of molecular electronics.

This work was supported by SPP~1243 of the Deutsche Forschungsgemeinschaft.


\end{document}